\documentclass[a4paper,11pt]{article}
\usepackage[latin1]{inputenc}
\usepackage{epsfig}
\usepackage[english]{babel}
\usepackage{url}
\usepackage[colorlinks=true,ps2pdf=true]{hyperref}

\def\rcs$#1${#1}
\newcommand{\version}[1]{\thanks{\rcs#1}}
\newcommand{\ie}{{i.e.},\ }		% i.e.
\newcommand{\eg}{{e.g.},\ }		% e.g.

\title{Increased security through open source
\version{$Id: oss-acm.tex,v 1.16 2005/04/26 17:44:29 jhh Exp $}}
\author{Jaap-Henk Hoepman, Bart Jacobs\\
  Security of Systems (SoS) group\\
  Institute for Computing and Information Sciences\\
  Radboud University Nijmegen\\
  P.O. Box 9010, 6500 GL \ Nijmegen, 
  the Netherlands\\ \texttt{\{jhh,bart\}@cs.ru.nl}}

\begin{document}

\maketitle

\bibliographystyle{plain}

%REFS:

%http://slashdot.org/articles/04/02/14/0148248.shtml?tid=126&tid=172

%http://slashdot.org/askslashdot/00/06/21/1333222.shtml
%http://www.linuxforum.dk/2000/slides/GeneS/GeneSpafford.pdf
%http://www.techtv.com/screensavers/print/0,23102,3406300,00.html

\section{Introduction}

The last few years have shown a worldwide rise in the attention for,
and actual use of, open source software (OSS), most notably of the
operating system Linux and various applications running on top of it. 
Various major companies and governments are
adopting OSS. As a result, there are many publications concerning its
advantages and disadvantages. The ongoing discussions cover a wide
range of topics, such as Windows versus Linux, cost issues, intellectual
property rights, development methods, etc. Here we wish to focus on security
issues surrounding OSS. It has become a reasonably well-established
conviction within the computer security community that publishing
designs and protocols contributes to the security of systems built on them.
But should one go all the way and publish source code as well? That is
the fundamental question that we wish to address in this paper.

The following analogies may help to introduce the issues and controversies
surrounding the open source debate.
\begin{itemize}
\item
Would you, while travelling far from home, take medicines of an unknown brand
given to you by a self-proclaimed ``doctor'', without
documentation, and hence without (independent) assurance about the
nature and proper working of the ingredients?
\item
Who would you trust most? A locksmith who keeps the working of his locks
secret, so that thieves cannot exploit this knowledge? Or a locksmith
who publishes the workings of his locks, so that
everyone (including thieves) can judge how good/bad they are (so you
exclusively rely on the complexity of the keys for protection)?
\end{itemize}

In the remainder of this paper we will discuss the impact of open source on
both the security and transparency of a software system. We focus on the
more technical aspects of this issue (and refer to
Glass~\cite{glass2004economics} for a discussion of the economical perspective
of open source), combining and extending arguments developed over the years
\cite{witten2001opensourcesecurity}.
%,hissam2002trustopensource}.
We stress that our discussion of the problem only applies to software for
general purpose computing systems. For embedded systems, where the software 
usually cannot easily be patched or upgraded, different considerations may
apply. 

\subsection{Security through obscurity: design vs. implementation}

% difference between desing and implementation

Through the centuries, secrecy was the predominant methodology
surrounding the design of any secure system. Security of military communication
systems, for example, was mostly based on the fact that only few people knew
how it worked, and not on any inherently secure method of
communication. Ciphers in those days were very insecure. 

In 1883, Auguste Kerckhoffs~\cite{kerckhoffs1883cryptographie} extensively
argued that any secure military system "... must not require secrecy and can be
stolen by the enemy without causing trouble". In the academic security
community Kerckhoffs' Principle is widely supported: in the design of a system,
\emph{security through obscurity} is considered bad practice,
for many reasons similar to the ones we will discuss later on.
This
point is starting to get across to industry as well, witnessed by the fact
that, for instance, the security of the third generation of cellular telephone
networks (UMTS) is based on open and published standards.

In Kerckhoffs' days there was hardly any difference between the
\emph{design} of a system and its actual \emph{implementation}. These days,
however, the
difference is huge: system designs are already very complex, and their
implementation is hard to get completely right. The question then arises
whether Kerckhoffs' Principle applies only to the design of a system, or also
to its implementation. In other words, should secure systems also be open
source, or not? 

There is no agreement on the answer to this question even in
the academic community~\cite{anderson2002boltzmann}.
According to us, the answer is: "absolutely!". In the
remainder of this paper we will argue why.

\subsection{Security, risk and exposure}

When discussing whether open source makes systems more secure, we have to be
precise about what we mean by that. In fact, for the purpose of this
discussion we need to distinguish between the
\emph{security} of a system, the \emph{exposure} of that system, and the
\emph{risk} associated with using that system. We define these terms next.

The ultimate decisive factor that determines whether a system is ``secure
enough'' is the \emph{risk} associated
with using that system. This risk is defined as a combination of the
likelihood of a successful attack on a system together with the damage
to the assets resulting from it.
%A less secure system that only protects assets of
%little value may be safer (\ie less risky) than a more secure system protecting
%valuable items. A risk analysis usually tries to estimate this risk.

The \emph{exposure} of a system completely ignores the damage that is
incurred by a successful attack, and is defined as just the likelihood of a
successful attack. This depends on several factors, like the number and
severity of vulnerabilities in the system, but also whether these
vulnerabilities are known to attackers, how hard it is to exploit such a
vulnerability, and whether the system is a high-profile target or not. 

Finally, we consider the \emph{security} of a system to be an objective measure
of the number of its vulnerabilities and their severity (\ie the privileges
obtained by exploiting the vulnerability). 

To summarise, exposure combines security with the likelihood of attack, and
risk combines exposure with the damage sustained by the attack.
We note that in other papers on
this and similar topics, security has been used to mean either security proper,
exposure or risk as defined above.
With these definitions in place, we see that
opening the source clearly does not change the security of a system
(simply because it doesn't introduce new bugs), while
the exposure is likely to increase in the short term
(because it makes the existing bugs more visible). The question is what
happens to the security and the exposure of an open source system in the long
run. 

\subsection{Open vs. closed source}

%- what do we mean by open source:
%  - NOT (necessarily) the development model (i.e. bazaar)

The increased attention paid to open source in the media and by society at
large has made open source an almost catch-all phrase. Here, we use it in its
original, rather specific, meaning. Open source software is software for
which the corresponding source (and all relevant documentation) is available
for inspection, use, modification and redistribution by the user\footnote{%
	See \url{http://www.opensource.org/}.
}.
We do not distinguish between any
kind of development methodology (\eg the Cathedral or the 
Bazaar\cite{raymond2000cathedral}). Neither do
we care about the pricing model
(freeware, shareware, etc.). We do assume however that users (in principle) are
allowed and able to rebuild the system from the (modified) sources, and that
they have access to the proper tools to do so. 

In some cases, allowing
the user to redistribute the modified sources (in full, or through patches)
is also necessary (\eg Free Software and the GNU Public License\footnote{%
	\url{http://www.gnu.org/copyleft/gpl.html}
}).
Most of our arguments also hold for \emph{source available} 
software, where the license does not allow redistribution of the (modified)
source.

\section{Open source necessary for security}

We believe that using open source software is a necessary requirement to build
systems that are more secure. 
Our main argument is that opening the source allows independent assessment of
the exposure of a system and 
the risk associated with using the system, makes patching bugs easier and
more likely, and forces software developers to spend more effort on the quality
of their code. The remainder of this paper is devoted to arguing our case in
detail. 

We will first review arguments in favour of keeping the source closed, and
then discuss reasons why open source does (in the long run) increase
security. As noted in the introduction, there is a distinction between making
the design of a system public and also making its implementation public. We
focus on the latter case, but note that most (but not all) of these arguments
also apply to the question whether the design should be open or not.

\subsection{Keep the source closed: arguments against open source}

First of all, keeping the source closed prevents the attacker from having easy
access to information that may be helpful to successfully launch an
attack~\cite{Bro02}. 
Opening the source gives the attacker a wealth of information to search
for vulnerabilities and/or bugs, like potential buffer overflows, and thus
increases the exposure of the system.

Also, there is a huge difference between openness of the design and openness of
the source. Openness of the design may reveal logical errors in the security in
the worst case. With proper review, these errors can and usually are found. For
source code, this is not, or at least not completely, the case. In the
foreseeable future, source code will continue to contain bugs, no matter how
hard we look, test or verify. 

Moreover, opening the source gives unfair advantage to the attacker. The
attacker needs to find but \emph{one} vulnerability to successfully attack the
system.  The defender needs to patch \emph{all} vulnerabilities to protect
himself completely. This is considered an uneven battle.

Fourth, there is no direct guarantee that the binary object code running in
the computer corresponds to the source code that has been 
evaluated~\cite{thompson1984trustingtrust}. People
unable or unwilling 
to compile from source must rely on a trusted third party to vouch for
this\footnote{%
	Or could use tools like \texttt{systrace} to confine untrusted 
	object code, and to enforce a security 
	policy nevertheless~\cite{provos2003systrace}.
}.

Also, making the source public does not guarantee that any qualified person
will actually look at the source and evaluate (let alone improve) it. There are
many open source projects that, after a brief flurry of activity, are only
marginally maintained and quickly sink into oblivion.
The attackers, on the other hand, most surely \emph{will} scrutinise the
source. 

In bazaar style open source projects, back-doors may be sneaked into the source
by hackers posing as trustful contributors. That this is not an idle threat
became clear in November 2003, when Linux kernel developers discovered a
back-door in a harmlessly looking error-checking feature added to a
system call\footnote{%
	\url{http://www.securityfocus.com/news/7388}.
}. 

%Gene Spafford: "It's who writes it and whether it's planned [that makes a
%difference], not who looks at the code." 
%http://www.eweek.com/article2/0,3959,562220,00.asp

Finally, and more generally, the quality of a piece of software (and patches to
it) depend 
on the skills of the programmers working on it~\cite{neumann2003securityredux}.
For many open source projects
there is no a priori selection of programmers based on their skill. Usually any
help is appreciated, and there is only rudimentary quality control.

%"Unless there's a great deal of discipline underlying the development, there's
%no difference in the security [of proprietary and open-source software]. Open
%source is not inherently more secure," said Peter Neumann,  
%http://www.eweek.com/article2/0,3959,562220,00.asp

\subsection{Closed is not so closed: arguments against closed source}

Let us first review the arguments put forward against open
source in the previous section. 
The last two arguments against open source are actually aimed at the
development methodology instead. The systems developed in that manner would
also be more insecure if they were closed source. We assume a minimal standard
of proper coding practices, project management, change control and quality
control. 
In fact, one of our main points is that by opening up the source, software
projects cannot get away with poor project management and poor
quality control so easily.

Now turning to the first argument against closed source, we note that keeping
the source 
closed for a long time appears to be hard~\cite{mercuri2003obscurity}.
Last year, source code for certain
types of voting machines manufactured by Diebold were distributed on the
Internet, and subsequent research on that source code revealed horrible
programming errors and security vulnerabilities~\cite{kohno2004analysis}. 
Recently even parts of the source to Microsoft Windows NT became public. Within
days the first exploit based on this source code was published.
The Diebold case also revealed how bad coding standards of current closed 
source systems can be, and how they lead to awfully insecure systems.

Even if the source remains closed, vulnerabilities of such closed source
systems will eventually be found and become known to a larger public after a
while.  Vulnerabilities in existing closed source software are announced on a
daily basis. 
In fact, tools like debuggers and disassemblers allow attackers to find
vulnerabilities in applications without access to the source relatively
quickly.
%(the lock story of rubin? and the fact that the "profession" knew about this
%all along.)
Moreover, not all vulnerabilities that are discovered will be published. 
Their discoverers may keep them secret to avoid patches for them, allowing 
use of the vulnerability to exploit systems for a prolonged period of time.
We see that while the perceived exposure of a closed source system may appear
to be low, the actual exposure eventually becomes much higher (approaching
the exposure that would exist initially if the system were completely open
source). 

Even worse, only the producer of closed source software can release patches
for any vulnerabilities that are found. Many of those patches are released
weeks or months after the vulnerability is discovered, if at all. The latter
case occurs for instance with legacy software for which the company producing
it no longer exists (or refuses to give support for it after a while, \eg
Microsoft Windows NT Server 4.0 or Netscape Calendar).
The consequence is that systems stay exposed
longer, increasing the risk of using that system.

We see that keeping the source closed actually hurts the defender much
more than the attacker: while a determined attacker can still discover
weaknesses 
easily, the defender is prevented from patching them.

%- Closed source systems are generally not improved and tested incrementally.
%(Q: what is Microsoft's Windows/Office update release frequency these days?)

Finally, closed source software severely limits the user of such software 
to evaluate its security for or by himself. The situation improves if
at least the design of the system is open. If the system is evaluated by an
independent party according to some generally accepted methodology (like the
Common Criteria), this gives the user another basis for trusting the
security of the software. However such evaluations are rare (because they are
expensive), and usually limited to certain restricted usage scenarios or
parameter settings that may not correspond to the actual operating environment
of a particular user. Moreover, 
such evaluations apply only to a specific version of the software: 
new versions
need to be reevaluated.

%"I don't think it's a good idea to have one rule as to whether code should be
%open. If Microsoft opened the [Internet Explorer] code now, it would probably
%be very bad because it's full of all kinds of bugs. But if it had been open
%from the start, that would have been good," said Avi Rubin, 
%http://www.eweek.com/article2/0,3959,562220,00.asp

%Closure failures: GSM, CSS (but this is about design, not open \emph{source})

\subsection{The way forward: arguments supporting open source}

We see that the arguments against "security through obscurity" generally apply
to the implementation of a system as well. It is a widely held design
principle that the security of a system should only depend on the secrecy of
the (user-specific) keys, on the ground that all other information of the
system is shared by many other people and therefore will become public as a
matter of course.

Moreover, open source enables users to evaluate the security by themselves,
or to hire a party of \emph{their} choice to evaluate the security for them.
Open source even enables several different and independent teams of people to
evaluate the security of the system, removing the 
dependence on a single party to decide in favour or against a certain system.
All this does not decrease the security or exposure of the system. However,
it does help to asses the real exposure of the system, closing the gap
between perceived and actual exposure.

Open source enables users to find bugs, and to patch these bugs themselves.
There is also a potential network effect: if users submit their patches to a
central repository, all other users can update their system to include this
patch, increasing their security too. Given that
different users are likely to find different bugs, many bugs are potentially
removed. This leads to more and faster patches, and hence more secure code
(this corresponds to "Linus's Law": "Given enough eyeballs, bugs are 
shallow"~\cite{raymond2000cathedral}).
Evidence suggests that patches for open source software are released almost
twice as fast as for closed source software, thus halving the vulnerability
period~\cite{witten2001opensourcesecurity}.
%\rem{aka the Delphi Effect (?)}

If a user is unable to patch a bug himself, open source at least enables him
to communicate about bugs with developers more efficiently (because both
can use the same frame of reference --- \ie the source code --- for 
communication~\cite{raymond2000cathedral}).

Also, open source software enables users to add extra security measures.
Several tools exist to enhance the security of existing
systems, provided the source is available~\cite{cowan2003softwaresecurity}.
These tools do not rely on static checking of the code. Instead, they
add generic runtime checks to the code to detect \eg buffer overflows or stack
frame corruptions. Moreover, open source software allows the user to limit the
complexity of the system (and thereby increasing its security) by removing
unneeded parts.

Finally, and importantly, open source forces developer communities
to be more careful, and
to use the best possible tools to secure their systems. It also forces
them to use clean coding styles ("sloppy" code is untrustworthy), and to
put more effort into quality control. 
Otherwise, companies and individual programmers alike will loose respect and
credibility. 
As a side effect, this will stimulate
research and development in new, improved tools for software development,
testing and evaluation and perhaps even verification.

\section{Conclusions}

We conclude that opening the source of existing systems will at first increase
their exposure, due to the fact that more information about vulnerabilities
becomes available to attackers.  However, this exposure (and the associated
risk of using the system) can now be determined publicly. With closed source
systems the perceived exposure may appear to be low, while the actual
exposure (due to increased knowledge of the attackers) may be much higher.
%(see also figure~\ref{fig-resilience}).

Moreover, because the source is open, all interested parties can assess the
exposure of a system, hunt for bugs and issue patches for them, or otherwise
increase the security of the system.  Security fixes will quickly be available,
so that the period of increased exposure is short.

In the long run, openness of the source will increase its security.
Sloppy code is visible to everyone, and questions even the overall quality
of it. Any available tools to validate the source will be used more often
by the
producers. If not, the users will do it themselves, afterwards. New, much more
advanced, tools will be developed to improve the security of software even
further.
Open source allows users to make a much more informed choice about the security
of a system, based on their own or on independent judgement.

It is our conviction that all these benefits outweigh the disadvantages of
a short period of increased exposure.

\bibliography{strings,oss-acm}

\end{document}